%% file: main.tex
\documentclass[sigconf]{acmart}

\AtBeginDocument{%
 \providecommand\BibTeX{{%
   \normalfont B\kern-0.5em{\scshape i\kern-0.25em b}\kern-0.8em\TeX}}}
    
\setcopyright{none} 
\renewcommand\footnotetextcopyrightpermission[1]{} 
  \acmConference[MSR 2022]{MSR '22: Proceedings of the 19th International Conference on Mining Software Repositories}{May 23–24, 2022}{Pittsburgh, PA, USA}
\acmBooktitle{MSR '22: Proceedings of the 19th International Conference on Mining Software Repositories, May 23–24, 2022, Pittsburgh, PA, USA}
\acmPrice{15.00}
\acmISBN{978-1-4503-XXXX-X/18/06}

\usepackage{xspace}

\usepackage{array}
\newcolumntype{L}{>{\arraybackslash}m{16cm}}
\newcolumntype{C}[1]{>{\centering\let\newline\\arraybackslash\hspace{0pt}}m{#1}}
\newcolumntype{R}[1]{>{\raggedleft\let\newline\\arraybackslash\hspace{0pt}}m{#1}}
\usepackage[numbered]{bookmark}
\usepackage{tcolorbox}
\usepackage{graphicx}
\usepackage{xcolor}
\usepackage{stfloats}
\usepackage[justification=centering]{caption}
\usepackage[title]{appendix}
\usepackage{dirtytalk}
\usepackage{amsmath}
\usepackage{pgfplots, pgfplotstable}
\usetikzlibrary{patterns}

\usepackage[english]{babel}
\usepackage{textcomp}
\usepackage{mathtools}
\usepackage{ltxtable}
\usepackage{algorithm}
\usepackage{algpseudocode}
\usepackage{textcomp}

\usepackage{tcolorbox}

\usepackage{pgf-pie}
\definecolor{wedge1}{RGB}{ 190  30  46}
\definecolor{wedge2}{RGB}{ 240  65  54}
\definecolor{wedge3}{RGB}{ 241  90  43}
\definecolor{wedge4}{RGB}{ 247 148  30}
\definecolor{wedge5}{RGB}{  43  56 144}
\definecolor{wedge6}{RGB}{  28 117 188}
\definecolor{wedge7}{RGB}{  40 170 225}
\definecolor{wedge8}{RGB}{ 119 179 225}
\definecolor{wedge9}{RGB}{ 181 212 239}
\definecolor{wedge10}{RGB}{  0 104  56}
\definecolor{wedge11}{RGB}{  0 148  69}
\definecolor{wedge12}{RGB}{ 57 181  74}
\definecolor{wedge13}{RGB}{141 199  63}
\definecolor{wedge14}{RGB}{215 244  34}
\definecolor{wedge15}{RGB}{249 237  50}
\definecolor{wedge16}{RGB}{248 241 148}
\definecolor{wedge17}{RGB}{242 245 205}
\definecolor{wedge18}{RGB}{123  82  49}
\definecolor{wedge19}{RGB}{104  73 158}
\definecolor{wedge20}{RGB}{102  45 145}
\definecolor{wedge21}{RGB}{148 149 151}
\definecolor{wedge22}{RGB}{ 204 50 153}
\definecolor{wedge23}{RGB}{ 79 47 79}
\definecolor{wedge24}{RGB}{ 173 234 234}
\definecolor{wedge25}{RGB}{ 216 191 216}
\definecolor{wedge26}{RGB}{  43  56 144}
\definecolor{wedge27}{RGB}{  40 170 225}
\definecolor{wedge28}{RGB}{ 119 179 225}
\definecolor{wedge29}{RGB}{ 181 212 239}
\definecolor{wedge30}{RGB}{  0 104  56}
\definecolor{wedge31}{RGB}{  0 148  69}
\definecolor{wedge32}{RGB}{ 57 181  74}

\usetikzlibrary{chains}

\pgfmathsetmacro\startAngle{90-3.6/2}
\pgfmathsetmacro\radius{+5}
\pgfmathsetmacro\maxLeg{+12}
\pgfmathsetmacro\legBound{+60}
\pgfmathsetmacro\legSpacing{2*\legBound/(\maxLeg-1)}

\usepackage{verbatim}
\pgfplotsset{compat=1.14}
\usetikzlibrary{patterns,}

\usepackage{graphicx}
\usepackage{tabularx,booktabs}
\usepackage{array}
\usepackage{bbding}
\usepackage{pifont}
\usepackage{wasysym}
\usepackage{caption}
\usepackage{dirtytalk}
\usepackage{subcaption}
\usepackage{tabularx}
\usepackage{sfmath}
\usepackage{array, booktabs, makecell}
\usepackage{color}
\usepackage{csquotes}
\usepackage{url}
\usepackage{comment}
\usepackage{tikz}
\usepackage{balance}
\usepackage{listings}
\usepackage{booktabs} 
\usepackage{soul} 
\usetikzlibrary{fit} 
\usetikzlibrary{positioning}
\usetikzlibrary{arrows}
\usetikzlibrary{shapes.multipart}
\usepackage{adjustbox}
\usepackage{float}
\usepackage{rotating}
\usepackage{nth}
\DeclareCaptionType{TextBox}
\usepackage{pstricks}
\usepackage{placeins}
\usepackage{afterpage}
\usepackage{multirow} 
\usepackage{textcomp}
\usepackage{multicol}
\usepackage{varwidth} 

\renewcommand{\shortauthors}{AlOmar et al.}

\author{Eman Abdullah AlOmar}
\affiliation{
    \institution{Rochester Institute of Technology}
    \city{Rochester, New York}
    \country{USA}
}
\email{eman.alomar@mail.rit.edu}

\author{Anthony Peruma}
\affiliation{
    \institution{Rochester Institute of Technology}
    \city{Rochester, New York}
    \country{USA}
}
\email{axp6201@rit.edu}

\author{Mohamed Wiem Mkaouer}
\affiliation{
    \institution{Rochester Institute of Technology}
    \city{Rochester, New York}
    \country{USA}
}
\email{mwmvse@rit.edu}

\author{Christian D. Newman}
\affiliation{
    \institution{Rochester Institute of Technology}
    \city{Rochester, New York}
    \country{USA}
}
\email{cnewman@se.rit.edu}

\author{Ali Ouni}
\affiliation{
    \institution{ETS Montreal, University of Quebec}
    \city{Montreal, Quebec}
    \country{Canada}
}
\email{ali.ouni@etsmtl.ca}

\begin{document}

\title{An Exploratory Study on Refactoring Documentation in Issues Handling}



\begin{abstract}
    Understanding the practice of refactoring documentation is of paramount importance in academia and industry. Issue tracking systems are used by most software projects enabling developers, quality assurance, managers, and users to submit feature requests and other tasks such as bug fixing and code review. Although recent studies explored how to document refactoring in commit messages, little is known about how developers describe their refactoring needs in issues. In this study, we aim at exploring developer-reported refactoring changes 
  in issues to better understand what developers consider to be problematic in their code and how they handle it. Our approach relies on text mining 45,477 refactoring-related issues and identifying refactoring patterns from a diverse corpus of 77 Java projects by investigating issues associated with 15,833 refactoring operations and developers' explicit refactoring intention. 
  Our results show that (1) developers mostly use move refactoring related terms/phrases to target refactoring-related issues; and (2) developers tend to explicitly mention the improvement of specific quality attributes and focus on duplicate code removal. 
     We envision our findings enabling tool builders to support developers with automated documentation of refactoring changes in issues. 
\end{abstract}




\maketitle

\input{Sections/01_Introduction}

\input{Sections/03_StudyDesign}
\input{Sections/04_Results}

\input{Sections/05_Discussion}

\input{Sections/06_Threats}
\input{Sections/07_Conclusion}

\bibliographystyle{ACM-Reference-Format}
\bibliography{sample-base}

\end{document}

%% file: Sections/01_Introduction.tex
\section{Introduction}
\label{Section:Introduction}

Code refactoring is a disciplined software engineering practice that is known as \say{the process of changing a software system in such a way that it does not alter the external behavior of the code yet improves its internal structure} \cite{Fowler:1999:RID:311424,alomar2021preserving}. Refactoring is commonly used in different development and maintenance tasks \cite{alomar2021ESWA}. 
 It supports developers in revolving submitted issues such as feature requests \cite{nyamawe2019automated} or bug reports \cite{di2020relationship}. 
  Issue tracking systems are used by most contemporary software projects enabling developers, quality assurance, managers, and users to submit feature or enhancement requests, as well as other tasks such as bug fixing and code review. 

Previous studies have focused on recommending refactorings through the detection of refactoring opportunities, either by identifying code anti-patterns that need correction \cite{brito2020refactoring,oizumi2020recommending,lenarduzzi2020sonarqube}, or by optimizing code quality metrics \cite{mkaouer2015many,anwer2017effect,sellittotoward}. Yet, recent studies have shown that there is a gap between automated refactoring tools, and what developers consider to a need-to-refactor situation in code \cite{cedrim2016does,alomar2019impact,fernandes2020refactoring}. To bridge this gap, it is important to understand what triggers developers to refactor their code, and what do developers care about when it comes to code improvement. Such information provides insights, to software practitioners and researchers, about the developer’s perception of refactoring. This can question whether developers do care about structural metrics and code smells when refactoring their code, or if there are other factors that are of direct influence on these non-functional changes.

In this paper, we focus on investigating issues that are written to express a need for refactoring. We extract patterns differentiating refactoring issues. These patterns represent what developers consider to be worth refactoring. We are also interested in the solutions, \textit{i.e.}, refactoring operations, being proposed as correction measures. Our investigation is driven by answering the following research questions:

\begin{itemize}
    \item \textbf{RQ$_1$: \textit{What textual patterns do developers use to describe their refactoring needs in issues?}} This RQ explores the existence of refactoring documentation in issues containing refactorings. This RQ aims to identify developers' common phrases when describing their refactoring problem/challenge.
    \item \textbf{RQ$_2$: \textit{What are the quality attributes developers care about when documenting in issues?}} 
    In this RQ, we investigate whether developers explicitly indicate the purpose of their refactoring activity applied in issues, \textit{e.g.,} improving structural metrics of fixing code smells.
\end{itemize}

The results of this exploratory study strengthen our understanding of what circumstances 
cause the need for refactorings. 
 Using the evolution history of 77 open-source projects exhibiting a total of 45,477 refactoring commits with issues, our study reveals that developers are mainly driven by reducing complexity and increasing comprehension and performance. While various studies associate refactoring tightly with fixing code smells, the only anti-pattern that was highlighted is duplicate code. Furthermore, various studies have shown that the \textit{rename} category is the most frequent in terms of refactoring operations (\textit{e.g.}, rename method, rename attribute, etc.) \cite{peruma2018empirical,negara2013comparative},  
  but our study reveals that developers tend also to discuss more complex refactorings, including refactorings dealing with \textit{extract} and \textit{move} code fragments, due to their impact on the design and code semantics preservation. We have also prepared a replication package of issues and their corresponding fixes (refactorings), to support the reproducibility and extension of our work \cite{ReplicationPackage}.

%% file: Sections/03_StudyDesign.tex
\section{Study Design}
\label{Section:Methodology}

\begin{figure*}[h]
\centering 
\includegraphics[width=\textwidth]{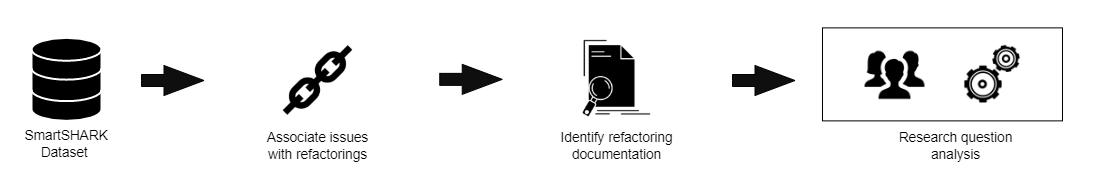}
\caption{Overview of our experiment design.} 
\label{fig:approach}
\end{figure*}

\input{Tables/Project_Overview}

Figure \ref{fig:approach} depicts a general overview of our experimental setup. In the following subsections, we elaborate on the activities involved in the process. We provide the dataset that we generate available in our replication package \cite{ReplicationPackage}.

\subsection{Source Dataset}
Our study utilizes the SmartSHARK MongoDB Release 2.1 dataset  \cite{msr2021dataset}. This dataset contains a wide range of information for 77 open-source Java projects, such as commit history, issues, refactorings, code metrics, mailing lists, and continuous integration data, among others. All 77 Java projects are part of the Apache ecosystem and utilize GitHub as their version control repository and JIRA for issue tracking. Furthermore, SmartSHARK utilizes RefDiff \cite{silva2017refdiff} and RefactoringMiner \cite{tsantalis2018accurate} to mine refactoring operations. Finally, the SmartSHARK dataset schema provides the necessary relationship attributes between data collections to join two or more related collection types.


\subsection{Refactoring Documentation Detection}

To identify refactoring documentation patterns in issues, we perform a series of manual and automated activities, as follows: 

\vspace{.1cm}
\noindent \textbf{Step \#1: 
Issues associated with a refactoring activity.} As our study focuses on issues and refactorings, our analysis is limited to issues where one or more refactoring operations were performed as part of the issue resolution. Hence, we first extracted all different refactorings from the source dataset. Next, we identify all commits containing the refactoring operations. Finally, we extracted issues that was addressed using the identified commits. 

\vspace{.1cm}
\noindent \textbf{Step \#2: Issues associated with developer intention about refactoring.} To ensure that the selected issues are about refactoring, we focused on a subset 
of 835 issues that reported developers' intention about the application of refactoring (\textit{i.e.}, having the keyword `\textit{refactor}'). The choice of `\textit{refactor}', besides being used by all related studies, is intuitively the first term to identify ideal refactoring-related issues \cite{murphy2008gathering,kim2014empirical,alomar2019can,alomar2021icse}. 
 Finally, to reduce the occurrence of false positives, we limited our analysis to only the occurrence of the term \textit{`refactor'} in the title of the issue as the title is a concise description of the problem \cite{peruma2022refactor, Rosen2016ESE}.




\vspace{.1cm}
\noindent \textbf{Step \#3: Annotation of issues.} When creating issues, developers use natural language to describe the issues. Hence, given the diverse nature of developers describing the problem, an automated approach to analyzing the issue text is not feasible. Therefore, we performed a manual analysis of the issue title and body to identify  refactoring documentation patterns. Next, we grouped this subset of issues based on specific patterns.  
 Further, to avoid redundancy of any pattern, we only considered one phrase if we found different patterns with the same meaning. For example, if we find patterns such as `simplifying the code', `code simplification', and `simplify code', we add only one of these similar phrases in the list of patterns. This enables having a list of the most insightful and unique patterns, and it also helps in making more concise patterns that are usable for readers. 
 

%% file: Tables/Project_Overview.tex
\begin{table}[h!]
\begin{center}
\caption{Dataset overview of refactoring-related issues.}

\label{Table:DATA_Overview}
\begin{adjustbox}{width=0.32\textwidth,center}
\begin{tabular}{lllll}\hline
\toprule
\bfseries Item & \bfseries Count \\
\midrule
Refactoring commits with issues & 45,477 \\
Refactoring commits with issues having keyword `\textit{refactor*}' in title & 835 \\
Refactoring operations associated with issues &  15,833 \\
\midrule
\multicolumn{2}{c}{\textbf{\textit{Issue Status}}}\\
\bfseries Item & \bfseries Count  \\
\midrule
Closed & 603 \\
In progress & 8 \\
Open & 28 \\
Resolved & 196 \\
\midrule
\multicolumn{2}{c}{\textbf{\textit{Issue Resolution}}}\\
\bfseries Item & \bfseries Count  \\
\midrule
Done & 4 \\
Fixed & 780 \\
Implemented & 8 \\
Resolved & 4 \\
Won't fix & 3 \\
\midrule
\multicolumn{2}{c}{\textbf{\textit{Issue Types}}}\\
\bfseries Item & \bfseries Count  \\
\midrule
Bug & 95 \\
Improvement & 390 \\
New Feature & 2\\
Story & 1\\
Sub-task & 71 \\
Task & 274 \\
Test & 2 \\
\bottomrule
\end{tabular}
\end{adjustbox}
\end{center}
\vspace{-.5cm}
\end{table}

%% file: Sections/04_Results.tex
\section{Experimental Results}
\label{Section:Result}

\subsection{RQ$_1$: What textual patterns do developers use to describe their refactoring needs in issues?}
\noindent\textbf{Methodology.} To identify refactoring documentation patterns, we manually inspect a subset of issues. 
These patterns are represented in the form of a keyword or phrase that frequently occurs in the issues associated with refactoring-related commits. 
\input{Tables/RQ1_Generic_Patterns}

\begin{figure}[H]
\centering 
\includegraphics[width=0.6\columnwidth]{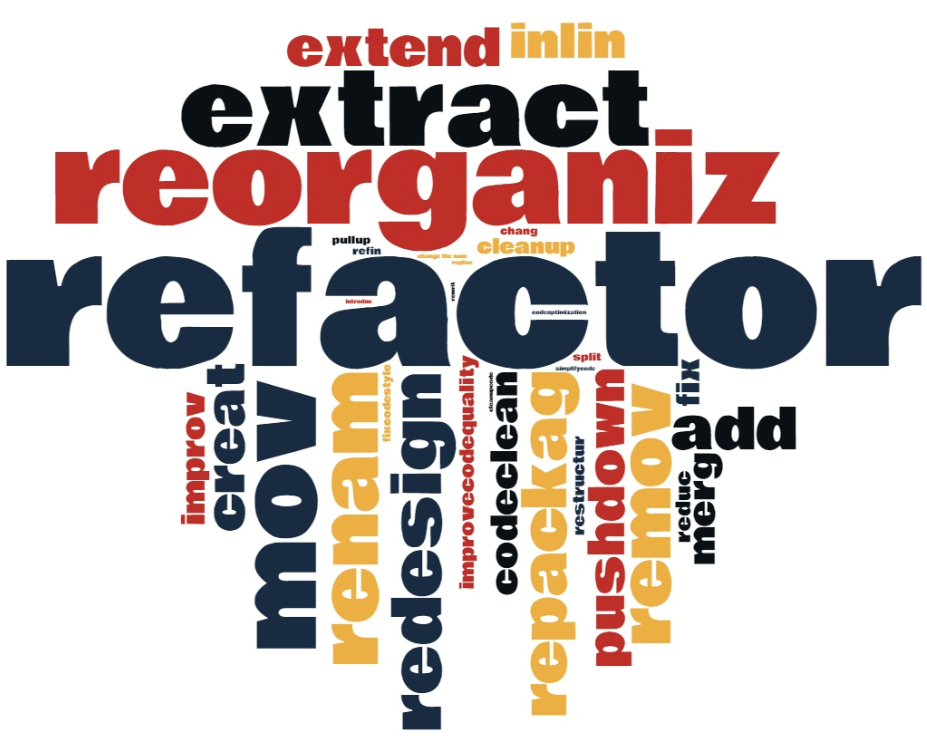}
\caption{Popular refactoring textual patterns in issues.}
\label{fig:Top Keywords}
\end{figure}

\noindent\textbf{Results.} Our in-depth inspection of the issues results in a list of 33 refactoring documentation patterns, as shown in Table~\ref{Table:GeneralPatterns}. Our findings show that the names of refactoring operations (\textit{e.g.}, `extract*', `mov*', `renam*') occur in the top frequently occurring patterns, and these patterns are mainly linked to code elements at different levels of granularity such as classes, methods, and variables.  These specific terms are well-known software refactoring operations and indicate developers' knowledge of the catalog of refactoring operations. We also observe that the top-ranked refactoring operation-related keywords include `extract*' and  `mov*'. `pull up' and `push down' operations are among the least discussed refactoring operations (similar to findings in \cite{Silva:2016:WWR:2950290.2950305,peruma2022refactor}). Moreover, we observe the occurrences of issue-fixing specific terms such as `fix*', `remov*', and `reduc*'. Next, we examine the most common keywords that developers use when expressing refactoring documentation in issues. Figure \ref{fig:Top Keywords} shows the top keywords used to identify refactoring documentations across the examined projects, that are ranked according to their number of occurrences. 
\input{Charts/RefactoringOperations}
\input{Tables/RQ2_Specific_Patterns}

\textcolor{black}{To better understand the nature of refactoring documentation, we have classified the associated refactoring operations
into 5 classes, namely, `changing', `extracting', `inlining', `moving', and `renaming'. Depicted in Figure \ref{fig:operation_clustered}, we cluster these operations associated with issues using a list of refactoring
keywords defined in a previous work \cite{alomar2021documentation}. The changing of the types belongs to the `changing' class, whereas the extraction of classes and methods are included in the `extracting' class. As for, `moving', it gathers all the movement of code elements, \textit{e.g.},
moving methods, or pushing code elements across hierarchies. Merging-related activities are included in the `inlining' class. Finally, the `renaming' class contains all refactorings that rename a given code element such as a class, a package or an attribute. As shown in Figure \ref{fig:operation_clustered}, the `extracting' operations are highly documented in issues across the projects,
and it reached the percentage of 37.2\%, higher than `moving', `renaming', and `changing' whose percentage is respectively 24.6\%, 21.5\%, and 14.7\%. The `inlining' operations, however, is the least documented refactoring which had a
ratio of only 2 \%.}

\subsection{RQ$_2$: What are the quality attributes developers care about when documenting in issues?}

\noindent\textbf{Methodology.} After identifying the different refactoring documentation patterns, we identify and categorize the patterns 
into three main categories (similar to \cite{alomar2019can,alomar2021ESWA,alomar2021toward,alomar2021behind}): (1) internal quality attributes, (2) external quality attributes, and (3) code smells. 

\noindent\textbf{Results.} Table \ref{Table:TopSpecificKeywords} provides the list of refactoring documentation patterns, ranked based on their frequency, we identify in refactoring-related issues. We observe that developers frequently mention key internal quality attributes (such as inheritance, complexity, etc.), a wide range of external quality attributes (such as readability and performance), and code duplication code smell that might impact code quality. To improve the internal design, the system structure optimization regarding its complexity and design size seems to be the dominant focus that is consistently mentioned in issues (0.26\% and 0.25\%, respectively). Concerning external quality attribute-related issues, we observe the mention of refactorings to enhance nonfunctional attributes. Patterns such as `readability', `performance', and `usability' represent the developers' main focus, with 0.23\%, 0.11\%, and 0.08\%, respectively. Finally, for code smell-focused refactoring issues, duplicate code represents the most popular anti-pattern developers intend to refactor (0.73\%).

%% file: Tables/RQ1_Generic_Patterns.tex
\begin{table}
\centering
\caption{List of refactoring documentation in issues (`*' captures the extension of the keyword) 
.}
\label{Table:GeneralPatterns}
\small
\begin{adjustbox}{width=0.8\columnwidth,totalheight=\textwidth,keepaspectratio}

\begin{tabular}{llllll}
\toprule
\textbf{Patterns}\\
\midrule
 Add* & Chang* &   Chang* the name \\
 Clean* up code &  Cleanup  & Code clean*  \\ 
Code optimization & Creat*  & Extend*    \\
Extract*  &  Fix*  & Fix* code style \\
Improv*  & Improv* code quality   & Inlin*  \\
  Introduc*  &   Merg* &  Mov* \\
 Pull* up &  Push* down & Repackag* \\
 Redesign* &   Reduc* &  Refactor* \\ 
 Refin* &  Remov*  & Renam* \\
Reorganiz* &  Replac* &   Restructur*  \\
 Rewrit* &  Simplify* code &  Split* \\
 \hline
       
\end{tabular}
\end{adjustbox}
\vspace{-.4cm}
\end{table}

%% file: Charts/RefactoringOperations.tex
\begin{figure}
\centering 
\begin{tikzpicture}
\begin{scope}[scale=0.6]
\pie[rotate = 180,pos ={0,0},text=inside,outside under=55,no number]{37.2/Extract\and37.2\%, 24.6/Move\and24.6\%, 21.5/Rename\and21.5\%,14.7/Change\and14.7\%,2.0/Inline\and2.0\%}
\end{scope}
\end{tikzpicture}
\caption{Percentage of refactorings, clustered by operation class.}
\label{fig:operation_clustered}
\end{figure}

%% file: Tables/RQ2_Specific_Patterns.tex
\begin{table}[H]
\fontsize{6.8}{6.8}\selectfont
\begin{center}
\caption{Summary of refactoring patterns, clustered by refactoring related categories.
} 
\label{Table:TopSpecificKeywords}
\begin{tabular}{lll}

\toprule
\textbf{Internal QA (\%)} &
\textbf{External QA  (\%)} &
\textbf{Code Smell  (\%)} \\
         \midrule 
    
  Complexity (0.26 \%)   & Readability (0.23 \%) &  Duplicate code (0.73 \%) \\ 
  Design Size (0.25 \%)    & Performance (0.11 \%)   &    \\ 
  Encapsulation (0.18 \%)     &   Usability (0.08 \%)  &  \\ 
  Dependency (0.16 \%)    & Extensibility (0.07 \%)   &   \\ 
  Inheritance (0.08 \%)    & Compatibility (0.06 \%)  &  \\
  Coupling (0.02 \%)    & Accuracy (0.05 \%)  &  \\
  Abstraction (0.02 \%)   & Modularity (0.05 \%)   & \\
    & Flexibility (0.04 \%)    & \\
     & Understandability (0.04 \%)  &  \\
      & Reusability (0.04 \%)  & \\ 
     & Testability (0.03 \%)   & \\ 
     & Maintainability (0.03 \%)   &  \\
    & Manageability (0.02 \%) & \\ 
       &  Stability (0.006 \%) & \\
    & Accessibility (0.01 \%)   & \\
    & Configurability (0.01 \%)   &  \\
    & Robustness (0.006 \%)  & \\
    & Repeatability (0.006 \%)  & \\
    &  Effectiveness (0.006 \%)  & \\
        \bottomrule
\end{tabular} 
\end{center}
\vspace{-.3cm}
\end{table}

%% file: Sections/05_Discussion.tex
\section{Discussion}
\label{Section:Discussion}


Our research aims to explore refactoring documentation in issues to provide future research directions that support developers in understanding refactoring applied in issues.

\textbf{RQ$_1$} indicates that developers tend to use a variety of textual patterns to document their refactorings in issues. These patterns can provide either a  (1) generic description of problems developers encounter or (2) a specific refactoring operation name following Fowler's names \cite{Fowler:1999:RID:311424}. Although previous studies show that rename refactorings are a common type of refactoring, \textit{e.g.}, \cite{peruma2018empirical}, we notice that `mov*' and `extract*' are the topmost documented refactorings in issues. This can be explained by
the fact that developers tend to make many design improvement decisions that include remodularizing packages by moving classes, reducing class-level coupling, and
increasing cohesion by moving methods. Additionally, developers might use similar terminology when performing move-related refactoring operations, \textit{i.e.}, \textit{Extract/Inline/Pull-up/Push-down} \cite{alomar2021documentation}. \textcolor{black}{As shown in Figure \ref{fig:operation_clustered}, `extracting' is the most documented refactorings. An interpretation for this comes from the nature of the debugging process that may include the separation of concerns which helps in reducing
the core complexity of a larger module and reduce its proneness to errors \cite{tsantalis2011identification}}. In other words, developers tend to discuss more complex refactorings in issues, including refactorings from the \textit{extract} and \textit{move} categories, due to their impact on the design and code semantic preservation. This information can provide valuable references for refactoring documentation practice in issues. For example, whether refactoring-related issue descriptions have the relevant information is a critical indicator for reproducing refactoring-related issues. 

From \textbf{RQ$_2$}, we observe that developers discuss quality concerns when documenting refactorings in issues that can be related to: (1) internal quality attributes, (2) external quality attributes, or (3) code smells. When analyzing these quality concerns per issue types reported in Table \ref{Table:DATA_Overview}, we notice that complexity and duplicate code are mostly documented with issue type named `bug', whereas the duplicate code and readability were the popular sub-categories for refactoring-related issues type named `improvement'. For instance, the  developer discussed fixing design issues by putting common functionalities into a superclass to eliminate duplicate code, breaking up lengthier methods to make the code more readable, and avoiding nested complex data structure to reduce code complexity. Moreover, we observe that code smell is rarely documented in issues. As shown in Table \ref{Table:TopSpecificKeywords}, developers only focused on duplicate code removal. Conversely, developers tend to report a variety of external quality attributes, focusing mainly on improving \textit{readability} of the code. This corroborates the finding by Palomba et al. \cite{palomba2017exploratory},  where refactoring targeting program comprehension was mostly applied during bug fixing activities. 
 As developers discussed functional and non-functional aspects of source code, future research can further investigate the intent as to why and how developers perform refactoring in issues. With a better understanding of this phenomenon, researchers and tool builders can support developers with automatically documenting refactorings in issues. 


One of the main purposes of exploring refactoring documentation in issues is to better understand how developers cope with their software decay by extracting any refactoring strategies that can be associated with removing code smells \citep{tsantalis2008jdeodorant,bavota2013empirical}, or improving the design structural measurements \citep{mkaouer2014recommendation,bavota2014recommending}. However, these techniques only analyze the changes at the source code level, and provide the operations performed, without associating it with any textual description, which may infer the rationale behind the refactoring application. Our proposal, of textual patterns, is the first step towards complementing the existing effort in detecting refactorings, by augmenting it with any description that was intended to describe the refactoring activity. As previously shown in Tables \ref{Table:GeneralPatterns} and \ref{Table:TopSpecificKeywords}, developers tend to add a high-level description of their refactoring activity, and mention their intention behind refactoring (remove duplicate code, improve readability, etc.), along with mentioning the refactoring operations they apply (type migration, inline methods, etc.).

Overall, the documentation of refactoring in issues is an important research direction that requires further attention. It has been known that there is a general shortage of refactoring documentation, and there is no consensus about how refactoring should be documented, which makes it subjective and developer-specific. Lack of design documentation forced developers to rely on
the source code to identify design problems \cite{sousa2018identifying}. Moreover, the fine-grained description of refactoring can be time-consuming, as a typical description should contain an indication about the operations performed, refactored code elements, and a hint about the intention behind the refactoring. In addition, the developer specification can be ambiguous as it reflects the developer's understanding of what has been improved in the source code, which can be different in reality, as the developer may not necessarily adequately estimate the refactoring impact on the quality improvement.

%% file: Sections/06_Threats.tex
\vspace{-.2cm}
\section{Threats To Validity}
\label{Section:Threats}

The first threat relates to the analysis of open-source Java projects. Our results may not generalize to systems written in other languages. Another potential threat to validity relates to our findings regarding counting the reported quality attributes and code smells. Due to the large number of commit messages, we have not performed a manual validation to remove false positive commit messages. Thus, this may have an impact on our findings. Finally, we constructed our dataset by extracting issues containing the term `refactor' in the title. There is the possibility that we may have excluded synonymous terms/phrases. However, even though this approach reduces the number of issues in our dataset, it also decreases false-positives, and ensures that we analyze issues that are explicitly focused on refactorings.

%% file: Sections/07_Conclusion.tex
\section{Conclusion \& Future Work}
\label{Section:Conclusion}

In this study, we performed an exploratory study to understand how developers document refactorings in issues. Specifically, we identify refactoring terms/phrases patterns, study possible refactoring documentation types, and determine how many refactoring terms/phrases exist in issues. Our results show that (1) developers mostly use move refactoring related terms/phrases to target refactoring-related issues; and (2) developers tend to explicitly mention the improvement of specific quality attributes and focus on duplicate code removal. 
 We envision our findings enabling tool builders to support developers with automatically document refactoring in issues. Future work in this area includes investigating which refactoring operation is more problematic in issues.

%% file: main.bbl

\begin{thebibliography}{37}


\ifx \showCODEN    \undefined \def \showCODEN     #1{\unskip}     \fi
\ifx \showDOI      \undefined \def \showDOI       #1{#1}\fi
\ifx \showISBNx    \undefined \def \showISBNx     #1{\unskip}     \fi
\ifx \showISBNxiii \undefined \def \showISBNxiii  #1{\unskip}     \fi
\ifx \showISSN     \undefined \def \showISSN      #1{\unskip}     \fi
\ifx \showLCCN     \undefined \def \showLCCN      #1{\unskip}     \fi
\ifx \shownote     \undefined \def \shownote      #1{#1}          \fi
\ifx \showarticletitle \undefined \def \showarticletitle #1{#1}   \fi
\ifx \showURL      \undefined \def \showURL       {\relax}        \fi
\providecommand\bibfield[2]{#2}
\providecommand\bibinfo[2]{#2}
\providecommand\natexlab[1]{#1}
\providecommand\showeprint[2][]{arXiv:#2}

\bibitem[\protect\citeauthoryear{AlOmar}{AlOmar}{2022}]%
        {ReplicationPackage}
\bibfield{author}{\bibinfo{person}{AlOmar}.} \bibinfo{year}{2022}\natexlab{}.
\newblock \bibinfo{booktitle}{\emph{ReplicationPackage}}.
\newblock
\urldef\tempurl%
\url{https://smilevo.github.io/self-affirmed-refactoring/}
\showURL{%
\tempurl}


\bibitem[\protect\citeauthoryear{AlOmar, AlRubaye, Mkaouer, Ouni, and
  Kessentini}{AlOmar et~al\mbox{.}}{2021a}]%
        {alomar2021icse}
\bibfield{author}{\bibinfo{person}{Eman~Abdullah AlOmar},
  \bibinfo{person}{Hussein AlRubaye}, \bibinfo{person}{Mohamed~Wiem Mkaouer},
  \bibinfo{person}{Ali Ouni}, {and} \bibinfo{person}{Marouane Kessentini}.}
  \bibinfo{year}{2021}\natexlab{a}.
\newblock \showarticletitle{Refactoring Practices in the Context of Modern Code
  Review: An Industrial Case Study at Xerox}. In \bibinfo{booktitle}{\emph{2021
  IEEE/ACM 43rd International Conference on Software Engineering: Software
  Engineering in Practice (ICSE-SEIP)}}. IEEE, \bibinfo{pages}{348--357}.
\newblock


\bibitem[\protect\citeauthoryear{AlOmar, Liu, Addo, Mkaouer, Newman, Ouni, and
  Yu}{AlOmar et~al\mbox{.}}{2022}]%
        {alomar2021documentation}
\bibfield{author}{\bibinfo{person}{Eman~Abdullah AlOmar},
  \bibinfo{person}{Jiaqian Liu}, \bibinfo{person}{Kenneth Addo},
  \bibinfo{person}{Mohamed~Wiem Mkaouer}, \bibinfo{person}{Christian Newman},
  \bibinfo{person}{Ali Ouni}, {and} \bibinfo{person}{Zhe Yu}.}
  \bibinfo{year}{2022}\natexlab{}.
\newblock \showarticletitle{On the documentation of refactoring types}.
\newblock \bibinfo{journal}{\emph{Automated Software Engineering}}
  \bibinfo{volume}{29}, \bibinfo{number}{1} (\bibinfo{year}{2022}),
  \bibinfo{pages}{1--40}.
\newblock


\bibitem[\protect\citeauthoryear{AlOmar, Mkaouer, Newman, and Ouni}{AlOmar
  et~al\mbox{.}}{2021c}]%
        {alomar2021preserving}
\bibfield{author}{\bibinfo{person}{Eman~Abdullah AlOmar},
  \bibinfo{person}{Mohamed~Wiem Mkaouer}, \bibinfo{person}{Christian Newman},
  {and} \bibinfo{person}{Ali Ouni}.} \bibinfo{year}{2021}\natexlab{c}.
\newblock \showarticletitle{On preserving the behavior in software refactoring:
  A systematic mapping study}.
\newblock \bibinfo{journal}{\emph{Information and Software Technology}}
  (\bibinfo{year}{2021}), \bibinfo{pages}{106675}.
\newblock


\bibitem[\protect\citeauthoryear{AlOmar, Mkaouer, and Ouni}{AlOmar
  et~al\mbox{.}}{2019a}]%
        {alomar2019can}
\bibfield{author}{\bibinfo{person}{Eman~Abdullah AlOmar},
  \bibinfo{person}{Mohamed~Wiem Mkaouer}, {and} \bibinfo{person}{Ali Ouni}.}
  \bibinfo{year}{2019}\natexlab{a}.
\newblock \showarticletitle{Can refactoring be self-affirmed? an exploratory
  study on how developers document their refactoring activities in commit
  messages}. In \bibinfo{booktitle}{\emph{International Workshop on
  Refactoring-accepted. IEEE}}.
\newblock


\bibitem[\protect\citeauthoryear{AlOmar, Mkaouer, and Ouni}{AlOmar
  et~al\mbox{.}}{2021b}]%
        {alomar2021toward}
\bibfield{author}{\bibinfo{person}{Eman~Abdullah AlOmar},
  \bibinfo{person}{Mohamed~Wiem Mkaouer}, {and} \bibinfo{person}{Ali Ouni}.}
  \bibinfo{year}{2021}\natexlab{b}.
\newblock \showarticletitle{Toward the automatic classification of
  self-affirmed refactoring}.
\newblock \bibinfo{journal}{\emph{Journal of Systems and Software}}
  \bibinfo{volume}{171} (\bibinfo{year}{2021}), \bibinfo{pages}{110821}.
\newblock


\bibitem[\protect\citeauthoryear{AlOmar, Mkaouer, Ouni, and Kessentini}{AlOmar
  et~al\mbox{.}}{2019b}]%
        {alomar2019impact}
\bibfield{author}{\bibinfo{person}{Eman~Abdullah AlOmar},
  \bibinfo{person}{Mohamed~Wiem Mkaouer}, \bibinfo{person}{Ali Ouni}, {and}
  \bibinfo{person}{Marouane Kessentini}.} \bibinfo{year}{2019}\natexlab{b}.
\newblock \showarticletitle{On the impact of refactoring on the relationship
  between quality attributes and design metrics}. In
  \bibinfo{booktitle}{\emph{2019 ACM/IEEE International Symposium on Empirical
  Software Engineering and Measurement (ESEM)}}. IEEE, \bibinfo{pages}{1--11}.
\newblock


\bibitem[\protect\citeauthoryear{AlOmar, Peruma, Mkaouer, Newman, Ouni, and
  Kessentini}{AlOmar et~al\mbox{.}}{2021e}]%
        {alomar2021ESWA}
\bibfield{author}{\bibinfo{person}{Eman~Abdullah AlOmar},
  \bibinfo{person}{Anthony Peruma}, \bibinfo{person}{Mohamed~Wiem Mkaouer},
  \bibinfo{person}{Christian Newman}, \bibinfo{person}{Ali Ouni}, {and}
  \bibinfo{person}{Marouane Kessentini}.} \bibinfo{year}{2021}\natexlab{e}.
\newblock \showarticletitle{How we refactor and how we document it? On the use
  of supervised machine learning algorithms to classify refactoring
  documentation}.
\newblock \bibinfo{journal}{\emph{Expert Systems with Applications}}
  \bibinfo{volume}{167} (\bibinfo{year}{2021}), \bibinfo{pages}{114176}.
\newblock


\bibitem[\protect\citeauthoryear{AlOmar, Peruma, Mkaouer, Newman, and
  Ouni}{AlOmar et~al\mbox{.}}{2021d}]%
        {alomar2021behind}
\bibfield{author}{\bibinfo{person}{Eman~Abdullah AlOmar},
  \bibinfo{person}{Anthony Peruma}, \bibinfo{person}{Mohamed~Wiem Mkaouer},
  \bibinfo{person}{Christian~D Newman}, {and} \bibinfo{person}{Ali Ouni}.}
  \bibinfo{year}{2021}\natexlab{d}.
\newblock \showarticletitle{Behind the scenes: On the relationship between
  developer experience and refactoring}.
\newblock \bibinfo{journal}{\emph{Journal of Software: Evolution and Process}}
  (\bibinfo{year}{2021}), \bibinfo{pages}{e2395}.
\newblock


\bibitem[\protect\citeauthoryear{Anwer, Adbellatif, Alshayeb, and Anjum}{Anwer
  et~al\mbox{.}}{2017}]%
        {anwer2017effect}
\bibfield{author}{\bibinfo{person}{Sajid Anwer}, \bibinfo{person}{Ahmad
  Adbellatif}, \bibinfo{person}{Mohammad Alshayeb}, {and}
  \bibinfo{person}{Muhammad~Shakeel Anjum}.} \bibinfo{year}{2017}\natexlab{}.
\newblock \showarticletitle{Effect of coupling on software faults: An empirical
  study}. In \bibinfo{booktitle}{\emph{2017 International Conference on
  Communication, Computing and Digital Systems (C-CODE)}}. IEEE,
  \bibinfo{pages}{211--215}.
\newblock


\bibitem[\protect\citeauthoryear{Bavota, Dit, Oliveto, Di~Penta, Poshyvanyk,
  and De~Lucia}{Bavota et~al\mbox{.}}{2013}]%
        {bavota2013empirical}
\bibfield{author}{\bibinfo{person}{Gabriele Bavota}, \bibinfo{person}{Bogdan
  Dit}, \bibinfo{person}{Rocco Oliveto}, \bibinfo{person}{Massimiliano
  Di~Penta}, \bibinfo{person}{Denys Poshyvanyk}, {and} \bibinfo{person}{Andrea
  De~Lucia}.} \bibinfo{year}{2013}\natexlab{}.
\newblock \showarticletitle{An empirical study on the developers' perception of
  software coupling}. In \bibinfo{booktitle}{\emph{Proceedings of the 2013
  International Conference on Software Engineering}}. IEEE Press,
  \bibinfo{pages}{692--701}.
\newblock


\bibitem[\protect\citeauthoryear{Bavota, Panichella, Tsantalis, Di~Penta,
  Oliveto, and Canfora}{Bavota et~al\mbox{.}}{2014}]%
        {bavota2014recommending}
\bibfield{author}{\bibinfo{person}{Gabriele Bavota},
  \bibinfo{person}{Sebastiano Panichella}, \bibinfo{person}{Nikolaos
  Tsantalis}, \bibinfo{person}{Massimiliano Di~Penta}, \bibinfo{person}{Rocco
  Oliveto}, {and} \bibinfo{person}{Gerardo Canfora}.}
  \bibinfo{year}{2014}\natexlab{}.
\newblock \showarticletitle{Recommending refactorings based on team
  co-maintenance patterns}. In \bibinfo{booktitle}{\emph{Proceedings of the
  29th ACM/IEEE international conference on Automated software engineering}}.
  ACM, \bibinfo{pages}{337--342}.
\newblock


\bibitem[\protect\citeauthoryear{Brito, Hora, and Valente}{Brito
  et~al\mbox{.}}{2020}]%
        {brito2020refactoring}
\bibfield{author}{\bibinfo{person}{Aline Brito}, \bibinfo{person}{Andre Hora},
  {and} \bibinfo{person}{Marco~Tulio Valente}.}
  \bibinfo{year}{2020}\natexlab{}.
\newblock \showarticletitle{Refactoring Graphs: Assessing Refactoring over
  Time}.
\newblock \bibinfo{journal}{\emph{arXiv preprint arXiv:2003.04666}}
  (\bibinfo{year}{2020}).
\newblock


\bibitem[\protect\citeauthoryear{Cedrim, Sousa, Garcia, and Gheyi}{Cedrim
  et~al\mbox{.}}{2016}]%
        {cedrim2016does}
\bibfield{author}{\bibinfo{person}{Diego Cedrim}, \bibinfo{person}{Leonardo
  Sousa}, \bibinfo{person}{Alessandro Garcia}, {and} \bibinfo{person}{Rohit
  Gheyi}.} \bibinfo{year}{2016}\natexlab{}.
\newblock \showarticletitle{Does refactoring improve software structural
  quality? A longitudinal study of 25 projects}. In
  \bibinfo{booktitle}{\emph{Proceedings of the 30th Brazilian Symposium on
  Software Engineering}}. ACM, \bibinfo{pages}{73--82}.
\newblock


\bibitem[\protect\citeauthoryear{Di~Penta, Bavota, and Zampetti}{Di~Penta
  et~al\mbox{.}}{2020}]%
        {di2020relationship}
\bibfield{author}{\bibinfo{person}{Massimiliano Di~Penta},
  \bibinfo{person}{Gabriele Bavota}, {and} \bibinfo{person}{Fiorella
  Zampetti}.} \bibinfo{year}{2020}\natexlab{}.
\newblock \showarticletitle{On the relationship between refactoring actions and
  bugs: a differentiated replication}. In \bibinfo{booktitle}{\emph{Proceedings
  of the 28th ACM Joint Meeting on European Software Engineering Conference and
  Symposium on the Foundations of Software Engineering}}.
  \bibinfo{pages}{556--567}.
\newblock


\bibitem[\protect\citeauthoryear{Fernandes, Ch{\'a}vez, Garcia, Ferreira,
  Cedrim, Sousa, and Oizumi}{Fernandes et~al\mbox{.}}{2020}]%
        {fernandes2020refactoring}
\bibfield{author}{\bibinfo{person}{Eduardo Fernandes},
  \bibinfo{person}{Alexander Ch{\'a}vez}, \bibinfo{person}{Alessandro Garcia},
  \bibinfo{person}{Isabella Ferreira}, \bibinfo{person}{Diego Cedrim},
  \bibinfo{person}{Leonardo Sousa}, {and} \bibinfo{person}{Willian Oizumi}.}
  \bibinfo{year}{2020}\natexlab{}.
\newblock \showarticletitle{Refactoring effect on internal quality attributes:
  What haven’t they told you yet?}
\newblock \bibinfo{journal}{\emph{Information and Software Technology}}
  \bibinfo{volume}{126} (\bibinfo{year}{2020}), \bibinfo{pages}{106347}.
\newblock


\bibitem[\protect\citeauthoryear{Fowler, Beck, Brant, Opdyke, and
  Roberts}{Fowler et~al\mbox{.}}{1999}]%
        {Fowler:1999:RID:311424}
\bibfield{author}{\bibinfo{person}{Martin Fowler}, \bibinfo{person}{Kent Beck},
  \bibinfo{person}{John Brant}, \bibinfo{person}{William Opdyke}, {and}
  \bibinfo{person}{don Roberts}.} \bibinfo{year}{1999}\natexlab{}.
\newblock \bibinfo{booktitle}{\emph{Refactoring: Improving the Design of
  Existing Code}}.
\newblock \bibinfo{publisher}{Addison-Wesley Longman Publishing Co., Inc.},
  \bibinfo{address}{Boston, MA, USA}.
\newblock
\showISBNx{0-201-48567-2}
\urldef\tempurl%
\url{http://dl.acm.org/citation.cfm?id=311424}
\showURL{%
\tempurl}


\bibitem[\protect\citeauthoryear{Kim, Zimmermann, and Nagappan}{Kim
  et~al\mbox{.}}{2014}]%
        {kim2014empirical}
\bibfield{author}{\bibinfo{person}{Miryung Kim}, \bibinfo{person}{Thomas
  Zimmermann}, {and} \bibinfo{person}{Nachiappan Nagappan}.}
  \bibinfo{year}{2014}\natexlab{}.
\newblock \showarticletitle{An empirical study of refactoringchallenges and
  benefits at microsoft}.
\newblock \bibinfo{journal}{\emph{IEEE Transactions on Software Engineering}}
  \bibinfo{volume}{40}, \bibinfo{number}{7} (\bibinfo{year}{2014}),
  \bibinfo{pages}{633--649}.
\newblock


\bibitem[\protect\citeauthoryear{Lenarduzzi, Lomio, Huttunen, and
  Taibi}{Lenarduzzi et~al\mbox{.}}{2020}]%
        {lenarduzzi2020sonarqube}
\bibfield{author}{\bibinfo{person}{Valentina Lenarduzzi},
  \bibinfo{person}{Francesco Lomio}, \bibinfo{person}{Heikki Huttunen}, {and}
  \bibinfo{person}{Davide Taibi}.} \bibinfo{year}{2020}\natexlab{}.
\newblock \showarticletitle{Are SonarQube Rules Inducing Bugs?}. In
  \bibinfo{booktitle}{\emph{2020 IEEE 27th International Conference on Software
  Analysis, Evolution and Reengineering (SANER)}}. IEEE,
  \bibinfo{pages}{501--511}.
\newblock


\bibitem[\protect\citeauthoryear{Mkaouer, Kessentini, Bechikh, Deb, and
  {\'O}~Cinn{\'e}ide}{Mkaouer et~al\mbox{.}}{2014}]%
        {mkaouer2014recommendation}
\bibfield{author}{\bibinfo{person}{Mohamed~Wiem Mkaouer},
  \bibinfo{person}{Marouane Kessentini}, \bibinfo{person}{Slim Bechikh},
  \bibinfo{person}{Kalyanmoy Deb}, {and} \bibinfo{person}{Mel
  {\'O}~Cinn{\'e}ide}.} \bibinfo{year}{2014}\natexlab{}.
\newblock \showarticletitle{Recommendation system for software refactoring
  using innovization and interactive dynamic optimization}. In
  \bibinfo{booktitle}{\emph{Proceedings of the 29th ACM/IEEE international
  conference on Automated software engineering}}. ACM,
  \bibinfo{pages}{331--336}.
\newblock


\bibitem[\protect\citeauthoryear{Mkaouer, Kessentini, Shaout, Koligheu,
  Bechikh, Deb, and Ouni}{Mkaouer et~al\mbox{.}}{2015}]%
        {mkaouer2015many}
\bibfield{author}{\bibinfo{person}{Wiem Mkaouer}, \bibinfo{person}{Marouane
  Kessentini}, \bibinfo{person}{Adnan Shaout}, \bibinfo{person}{Patrice
  Koligheu}, \bibinfo{person}{Slim Bechikh}, \bibinfo{person}{Kalyanmoy Deb},
  {and} \bibinfo{person}{Ali Ouni}.} \bibinfo{year}{2015}\natexlab{}.
\newblock \showarticletitle{Many-objective software remodularization using
  NSGA-III}.
\newblock \bibinfo{journal}{\emph{ACM Transactions on Software Engineering and
  Methodology (TOSEM)}} \bibinfo{volume}{24}, \bibinfo{number}{3}
  (\bibinfo{year}{2015}), \bibinfo{pages}{1--45}.
\newblock


\bibitem[\protect\citeauthoryear{Murphy-Hill, Black, Dig, and
  Parnin}{Murphy-Hill et~al\mbox{.}}{2008}]%
        {murphy2008gathering}
\bibfield{author}{\bibinfo{person}{Emerson Murphy-Hill},
  \bibinfo{person}{Andrew~P Black}, \bibinfo{person}{Danny Dig}, {and}
  \bibinfo{person}{Chris Parnin}.} \bibinfo{year}{2008}\natexlab{}.
\newblock \showarticletitle{Gathering refactoring data: a comparison of four
  methods}. In \bibinfo{booktitle}{\emph{Proceedings of the 2nd Workshop on
  Refactoring Tools}}. \bibinfo{pages}{1--5}.
\newblock


\bibitem[\protect\citeauthoryear{Negara, Chen, Vakilian, Johnson, and
  Dig}{Negara et~al\mbox{.}}{2013}]%
        {negara2013comparative}
\bibfield{author}{\bibinfo{person}{Stas Negara}, \bibinfo{person}{Nicholas
  Chen}, \bibinfo{person}{Mohsen Vakilian}, \bibinfo{person}{Ralph~E Johnson},
  {and} \bibinfo{person}{Danny Dig}.} \bibinfo{year}{2013}\natexlab{}.
\newblock \showarticletitle{A comparative study of manual and automated
  refactorings}. In \bibinfo{booktitle}{\emph{European Conference on
  Object-Oriented Programming}}. Springer, \bibinfo{pages}{552--576}.
\newblock


\bibitem[\protect\citeauthoryear{Nyamawe, Liu, Niu, Umer, and Niu}{Nyamawe
  et~al\mbox{.}}{2019}]%
        {nyamawe2019automated}
\bibfield{author}{\bibinfo{person}{Ally~S Nyamawe}, \bibinfo{person}{Hui Liu},
  \bibinfo{person}{Nan Niu}, \bibinfo{person}{Qasim Umer}, {and}
  \bibinfo{person}{Zhendong Niu}.} \bibinfo{year}{2019}\natexlab{}.
\newblock \showarticletitle{Automated recommendation of software refactorings
  based on feature requests}. In \bibinfo{booktitle}{\emph{2019 IEEE 27th
  International Requirements Engineering Conference (RE)}}. IEEE,
  \bibinfo{pages}{187--198}.
\newblock


\bibitem[\protect\citeauthoryear{Oizumi, Bibiano, Cedrim, Oliveira, Sousa,
  Garcia, and Oliveira}{Oizumi et~al\mbox{.}}{2020}]%
        {oizumi2020recommending}
\bibfield{author}{\bibinfo{person}{Willian Oizumi}, \bibinfo{person}{Ana~C
  Bibiano}, \bibinfo{person}{Diego Cedrim}, \bibinfo{person}{Anderson
  Oliveira}, \bibinfo{person}{Leonardo Sousa}, \bibinfo{person}{Alessandro
  Garcia}, {and} \bibinfo{person}{Daniel Oliveira}.}
  \bibinfo{year}{2020}\natexlab{}.
\newblock \showarticletitle{Recommending Composite Refactorings for Smell
  Removal: Heuristics and Evaluation}. In \bibinfo{booktitle}{\emph{Proceedings
  of the 34th Brazilian Symposium on Software Engineering}}.
  \bibinfo{pages}{72--81}.
\newblock


\bibitem[\protect\citeauthoryear{Palomba, Zaidman, Oliveto, and
  De~Lucia}{Palomba et~al\mbox{.}}{2017}]%
        {palomba2017exploratory}
\bibfield{author}{\bibinfo{person}{Fabio Palomba}, \bibinfo{person}{Andy
  Zaidman}, \bibinfo{person}{Rocco Oliveto}, {and} \bibinfo{person}{Andrea
  De~Lucia}.} \bibinfo{year}{2017}\natexlab{}.
\newblock \showarticletitle{An exploratory study on the relationship between
  changes and refactoring}. In \bibinfo{booktitle}{\emph{2017 IEEE/ACM 25th
  International Conference on Program Comprehension (ICPC)}}. IEEE,
  \bibinfo{pages}{176--185}.
\newblock


\bibitem[\protect\citeauthoryear{Peruma, Mkaouer, Decker, and Newman}{Peruma
  et~al\mbox{.}}{2018}]%
        {peruma2018empirical}
\bibfield{author}{\bibinfo{person}{Anthony Peruma},
  \bibinfo{person}{Mohamed~Wiem Mkaouer}, \bibinfo{person}{Michael~J Decker},
  {and} \bibinfo{person}{Christian~D Newman}.} \bibinfo{year}{2018}\natexlab{}.
\newblock \showarticletitle{An empirical investigation of how and why
  developers rename identifiers}. In \bibinfo{booktitle}{\emph{Proceedings of
  the 2nd International Workshop on Refactoring}}. \bibinfo{pages}{26--33}.
\newblock


\bibitem[\protect\citeauthoryear{Peruma, Simmons, AlOmar, Newman, Mkaouer, and
  Ouni}{Peruma et~al\mbox{.}}{2022}]%
        {peruma2022refactor}
\bibfield{author}{\bibinfo{person}{Anthony Peruma}, \bibinfo{person}{Steven
  Simmons}, \bibinfo{person}{Eman~Abdullah AlOmar},
  \bibinfo{person}{Christian~D Newman}, \bibinfo{person}{Mohamed~Wiem Mkaouer},
  {and} \bibinfo{person}{Ali Ouni}.} \bibinfo{year}{2022}\natexlab{}.
\newblock \showarticletitle{How do I refactor this? An empirical study on
  refactoring trends and topics in Stack Overflow}.
\newblock \bibinfo{journal}{\emph{Empirical Software Engineering}}
  \bibinfo{volume}{27}, \bibinfo{number}{1} (\bibinfo{year}{2022}),
  \bibinfo{pages}{1--43}.
\newblock


\bibitem[\protect\citeauthoryear{Rosen and Shihab}{Rosen and Shihab}{2016}]%
        {Rosen2016ESE}
\bibfield{author}{\bibinfo{person}{Christoffer Rosen} {and}
  \bibinfo{person}{Emad Shihab}.} \bibinfo{year}{2016}\natexlab{}.
\newblock \showarticletitle{What are mobile developers asking about? A large
  scale study using stack overflow}.
\newblock \bibinfo{journal}{\emph{Empirical Software Engineering}}
  \bibinfo{volume}{21}, \bibinfo{number}{3} (\bibinfo{date}{01 Jun}
  \bibinfo{year}{2016}), \bibinfo{pages}{1192--1223}.
\newblock
\showISSN{1573-7616}
\urldef\tempurl%
\url{https://doi.org/10.1007/s10664-015-9379-3}
\showDOI{\tempurl}


\bibitem[\protect\citeauthoryear{Sellitto, Iannone, Codabux, Lenarduzzi,
  De~Lucia, Palomba, and Ferrucci}{Sellitto et~al\mbox{.}}{[n.\,d.]}]%
        {sellittotoward}
\bibfield{author}{\bibinfo{person}{Giulia Sellitto}, \bibinfo{person}{Emanuele
  Iannone}, \bibinfo{person}{Zadia Codabux}, \bibinfo{person}{Valentina
  Lenarduzzi}, \bibinfo{person}{Andrea De~Lucia}, \bibinfo{person}{Fabio
  Palomba}, {and} \bibinfo{person}{Filomena Ferrucci}.}
  \bibinfo{year}{[n.\,d.]}\natexlab{}.
\newblock \showarticletitle{Toward Understanding the Impact of Refactoring on
  Program Comprehension}.
\newblock  (\bibinfo{year}{[n.\,d.]}).
\newblock


\bibitem[\protect\citeauthoryear{Silva, Tsantalis, and Valente}{Silva
  et~al\mbox{.}}{2016}]%
        {Silva:2016:WWR:2950290.2950305}
\bibfield{author}{\bibinfo{person}{Danilo Silva}, \bibinfo{person}{Nikolaos
  Tsantalis}, {and} \bibinfo{person}{Marco~Tulio Valente}.}
  \bibinfo{year}{2016}\natexlab{}.
\newblock \showarticletitle{Why We Refactor? Confessions of GitHub
  Contributors}. In \bibinfo{booktitle}{\emph{Proceedings of the 2016 24th ACM
  SIGSOFT International Symposium on Foundations of Software Engineering}}
  (Seattle, WA, USA) \emph{(\bibinfo{series}{FSE 2016})}.
  \bibinfo{publisher}{ACM}, \bibinfo{address}{New York, NY, USA},
  \bibinfo{pages}{858--870}.
\newblock
\showISBNx{978-1-4503-4218-6}
\urldef\tempurl%
\url{https://doi.org/10.1145/2950290.2950305}
\showDOI{\tempurl}


\bibitem[\protect\citeauthoryear{Silva and Valente}{Silva and Valente}{2017}]%
        {silva2017refdiff}
\bibfield{author}{\bibinfo{person}{Danilo Silva} {and}
  \bibinfo{person}{Marco~Tulio Valente}.} \bibinfo{year}{2017}\natexlab{}.
\newblock \showarticletitle{Refdiff: detecting refactorings in version
  histories}. In \bibinfo{booktitle}{\emph{2017 IEEE/ACM 14th International
  Conference on Mining Software Repositories (MSR)}}. IEEE,
  \bibinfo{pages}{269--279}.
\newblock


\bibitem[\protect\citeauthoryear{Sousa, Oliveira, Oizumi, Barbosa, Garcia, Lee,
  Kalinowski, de~Mello, Fonseca, Oliveira, et~al\mbox{.}}{Sousa
  et~al\mbox{.}}{2018}]%
        {sousa2018identifying}
\bibfield{author}{\bibinfo{person}{Leonardo Sousa}, \bibinfo{person}{Anderson
  Oliveira}, \bibinfo{person}{Willian Oizumi}, \bibinfo{person}{Simone
  Barbosa}, \bibinfo{person}{Alessandro Garcia}, \bibinfo{person}{Jaejoon Lee},
  \bibinfo{person}{Marcos Kalinowski}, \bibinfo{person}{Rafael de Mello},
  \bibinfo{person}{Baldoino Fonseca}, \bibinfo{person}{Roberto Oliveira},
  {et~al\mbox{.}}} \bibinfo{year}{2018}\natexlab{}.
\newblock \showarticletitle{Identifying design problems in the source code: A
  grounded theory}. In \bibinfo{booktitle}{\emph{Proceedings of the 40th
  International Conference on Software Engineering}}.
  \bibinfo{pages}{921--931}.
\newblock


\bibitem[\protect\citeauthoryear{Trautsch, Trautsch, and Herbold}{Trautsch
  et~al\mbox{.}}{2021}]%
        {msr2021dataset}
\bibfield{author}{\bibinfo{person}{Alexander Trautsch}, \bibinfo{person}{Fabian
  Trautsch}, {and} \bibinfo{person}{Steffen Herbold}.}
  \bibinfo{year}{2021}\natexlab{}.
\newblock \showarticletitle{MSR Mining Challenge: The SmartSHARK Repository
  Data}.
\newblock  (\bibinfo{year}{2021}).
\newblock


\bibitem[\protect\citeauthoryear{Tsantalis, Chaikalis, and
  Chatzigeorgiou}{Tsantalis et~al\mbox{.}}{2008}]%
        {tsantalis2008jdeodorant}
\bibfield{author}{\bibinfo{person}{Nikolaos Tsantalis},
  \bibinfo{person}{Theodoros Chaikalis}, {and} \bibinfo{person}{Alexander
  Chatzigeorgiou}.} \bibinfo{year}{2008}\natexlab{}.
\newblock \showarticletitle{JDeodorant: Identification and removal of
  type-checking bad smells}. In \bibinfo{booktitle}{\emph{2008 12th European
  Conference on Software Maintenance and Reengineering}}. IEEE,
  \bibinfo{pages}{329--331}.
\newblock


\bibitem[\protect\citeauthoryear{Tsantalis and Chatzigeorgiou}{Tsantalis and
  Chatzigeorgiou}{2011}]%
        {tsantalis2011identification}
\bibfield{author}{\bibinfo{person}{Nikolaos Tsantalis} {and}
  \bibinfo{person}{Alexander Chatzigeorgiou}.} \bibinfo{year}{2011}\natexlab{}.
\newblock \showarticletitle{Identification of extract method refactoring
  opportunities for the decomposition of methods}.
\newblock \bibinfo{journal}{\emph{Journal of Systems and Software}}
  \bibinfo{volume}{84}, \bibinfo{number}{10} (\bibinfo{year}{2011}),
  \bibinfo{pages}{1757--1782}.
\newblock


\bibitem[\protect\citeauthoryear{Tsantalis, Mansouri, Eshkevari, Mazinanian,
  and Dig}{Tsantalis et~al\mbox{.}}{2018}]%
        {tsantalis2018accurate}
\bibfield{author}{\bibinfo{person}{Nikolaos Tsantalis}, \bibinfo{person}{Matin
  Mansouri}, \bibinfo{person}{Laleh Eshkevari}, \bibinfo{person}{Davood
  Mazinanian}, {and} \bibinfo{person}{Danny Dig}.}
  \bibinfo{year}{2018}\natexlab{}.
\newblock \showarticletitle{Accurate and efficient refactoring detection in
  commit history}. In \bibinfo{booktitle}{\emph{2018 IEEE/ACM 40th
  International Conference on Software Engineering (ICSE)}}. IEEE,
  \bibinfo{pages}{483--494}.
\newblock


\end{thebibliography}
